\documentclass[aps,prb,twocolumn,amsmath, superscriptaddress, floatfix, showpacs]{revtex4}
\usepackage{graphicx}

\newcommand{\bv}[1]{{\bf #1}}

\begin{document}

\title{Hidden zero-temperature bicritical point in the two-dimensional anisotropic Heisenberg model: Monte Carlo simulations and proper finite-size scaling}
\author{Chenggang Zhou}
\affiliation{Center for Nanophase Materials Science and Computer Science and Mathematics Division, Oak Ridge National Laboratory, P. O. Box 2008, Oak Ridge Tennessee, 37831-6493 USA }
\affiliation{Center for Simulational Physics, University of Georgia, Athens Georgia, 30602 USA }

\author{D. P. Landau}
\affiliation{Center for Simulational Physics, University of Georgia, Athens Georgia, 30602 USA }
\date{\today}

\author{T. C. Schulthess}
\affiliation{Center for Nanophase Materials Science and Computer Science and Mathematics Division, Oak Ridge National Laboratory, P. O. Box 2008, Oak Ridge Tennessee, 37831-6493 USA }

\begin{abstract}
By considering the appropriate finite-size effect, we explain the connection between Monte Carlo
simulations of two-dimensional anisotropic Heisenberg antiferromagnet in a field and the early
renormalization group calculation for the bicritical point in $2+\epsilon$ dimensions.
We found  that the long length scale physics of the Monte Carlo simulations is indeed captured
by the anisotropic nonlinear $\sigma$ model. Our Monte Carlo data and analysis confirm
that the bicritical point in two dimensions is Heisenberg-like and occurs at $T=0$,
therefore the uncertainty in the phase diagram of this model is removed.
\end{abstract}
\pacs{ 
05.10.Cc 
05.70.Jk   
75.10.-b   
75.40.Mg  
}
\maketitle
\section{Introduction}
The two-dimensional anisotropic Heisenberg antiferromagnet has
recently been re-studied with extensive Monte Carlo
simulations.~\cite{Leidl05} Twenty five years after the first
attempt to delineate its phase diagram with Monte Carlo
simulations,~\cite{Landau81} many features of it have been
clarified. However, the nature of the ``spin-flop transition'' has
been an issue under debate, because the data from the simulations
have been inadequate to show unambiguously the thermodynamic limit
of the phase boundaries. There is no spin-flop transition in the
thermodynamic limit according to the renormalization group (RG)
calculations in $2+\epsilon$ dimensions.~\cite{Pelcovits76,
Nelson77,Kosterlitz78} Instead, there are two neighboring second
order phase boundaries and a disordered phase in between. By
tracing the phase boundaries from high to low temperatures, only
an upper bound for the bicritical temperature can be claimed.
Below this temperature, an apparent first order spin-flop
transition is indeed observed in both Monte Carlo simulations and
neutron scattering experiments on quasi-two-dimensional Heisenberg
antiferromagnetic systems with anisotropy.~\cite{Cowley93} It has
been generally agreed that the existing data are consistent with
the RG predictions, although some features near the spin-flop
line, e.g. the apparent hysteresis~\cite{Landau81} and the
unexpected crossing points in the Binder cumulant,~\cite{Leidl05}
have not been accounted for.

In this paper, we study the ``spin-flop transition'' of the XXZ model defined by
the Hamiltonian
\begin{equation}
\label{eq1}
{\cal H} = J \sum_{\left< i,j\right>}
\left[\Delta\left( S_i^xS_j^x +  S_i^yS_j^y \right) + S_i^zS_j^z\right]
-H\sum_i S_i^z.
\end{equation}
Here $S_i^x$, $S_i^y$, and $S_i^z$ are three components of a unit
vector located on site $i$ of a square lattice with periodic
boundary conditions in both directions. The anisotropy is given by
the parameter $\Delta$, and $H$ is the external magnetic field in
the $z$ direction.  By the term ``spin-flop transition'', we refer
to the boundary region between the antiferromagnetic (AF) phase
and the XY phase where two separate second order phase boundaries
cannot be identified in simulations of finite-size systems.  We
set $J=1$ and $\Delta=4/5$ for simplicity. Our results are
expected to be valid for $0<\Delta<1$, since no qualitatively
different behavior has been found for other values of $\Delta$,
and the phase diagram for $\Delta=4/5$ is most accurately
known.~\cite{Leidl05}

\begin{figure}
\includegraphics[width=\columnwidth]{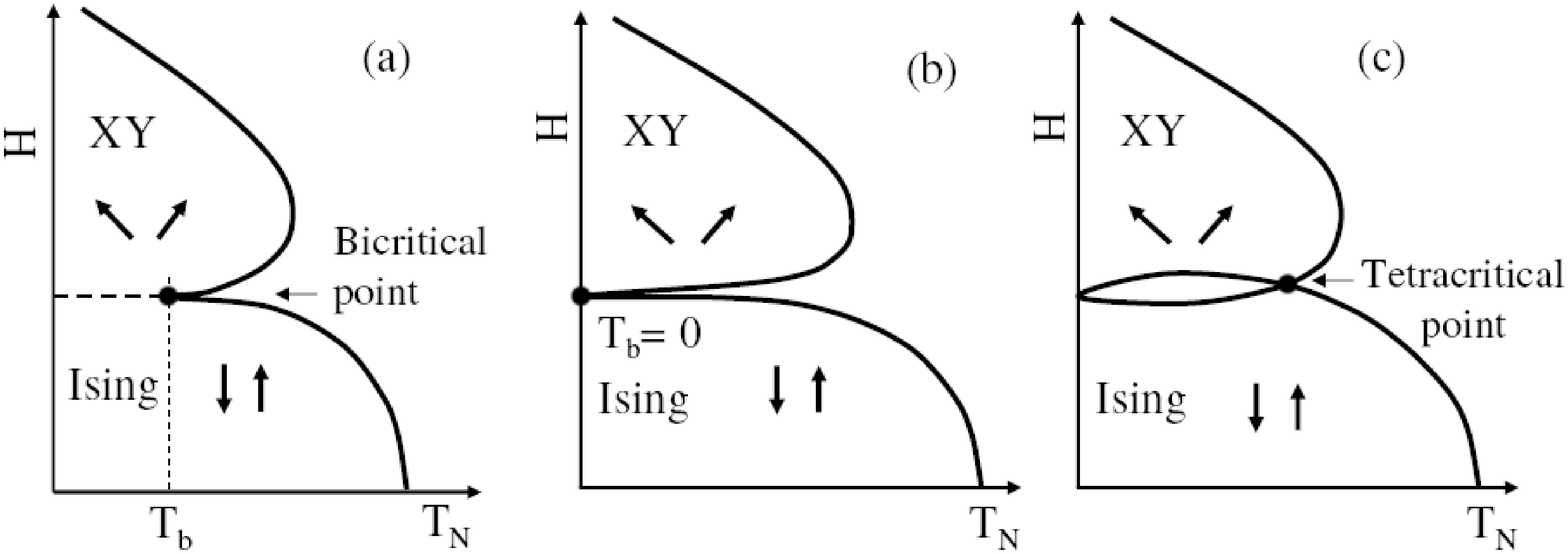}
\caption{Three candidates for the phase diagram. Solid lines are
second order phase boundaries, and the dashed line in (a) is a
first order phase boundary.} \label{fig0}
\end{figure}

At low temperatures and in a low magnetic field, systems described
by Eq.~(\ref{eq1}) exhibit an Ising-like AF phase, where the order
parameter, i.e. the staggered magnetization in the $z$ direction,
has a finite value. Above a sufficiently large magnetic field,
which is a function of temperature, the AF phase is replaced by an
XY phase, where the order parameter, i.e. the in-plane staggered
magnetization, has a finite value in the finite-size system due to the
power-law decay of its correlation function. As pointed out in
Ref.~\onlinecite{Leidl05}, there are three different possible
scenarios for this spin-flop transition: (a) It is a first order
phase transition at low temperature; the first order phase
boundary and the second order phase boundaries of the XY phase and
the AF phase meet at a bicritical point. (b) The bicritical point
appears at zero temperature with a very narrow disordered phase
separating the XY and AF phases. (c) A ``biconical phase'', in
which both order parameters are nonzero, separates the XY and the
AF phases. These three scenarios are shown in Fig.~\ref{fig0}
schematically.

Both RG calculations~\cite{Kosterlitz76} and the Monte Carlo
simulation~\cite{Landau78} have shown that (a) is realized in
three dimensions, a scenario which is consistent with the finite
critical temperature of the three dimensional isotropic Heisenberg
model. In two dimensions,  as a result of the Mermin-Wagner
theorem,~\cite{Mermin66} such a bicritical point can only be at
zero temperature, and the same prediction comes from RG
calculations with the non-linear $\sigma$
model.~\cite{Polyakov75,Pelcovits76,Zinn-Justin76,Brezin76,Nelson77,Kosterlitz78}
Because of the limited computer power that was then available,
early Monte Carlo simulations with the Hamiltonian of
Eq.~(\ref{eq1}) did not favor any of the above three
scenarios.~\cite{Landau81} Recently, Ref.~\onlinecite{Leidl05} has
found Ising-like scaling behavior in the specific heat and
susceptibility on a spin-flop line at $T/J = 0.33$ for
$\Delta=4/5$, which gives an upper bound in temperature for the
bicritical point. We have reproduced these scaling behaviors in
our simulations; however, our simulations at lower temperatures do
not display such Ising-like scaling behavior, as we will show in
this paper. Scenario (c) is realized in systems with random
fields~\cite{Aharony78} or spins with more than four components,
e.g. the SO(5) theory.~\cite{Hu01} It is unlikely to be the case
based on current and previous Monte Carlo simulations.

The main issue to address in this paper is how to compare our extensive data from
Monte Carlo simulations of limited system sizes with the RG calculations.
If they are consistent with each other, we will have a bicritical point at zero
temperature in the thermodynamic limit. Otherwise, we are forced to accept the
apparent first order phase transition that
we have seen at low temperatures.  Because the correlation length
is expected to diverge as $\exp(4\pi/T)$ at low temperatures, our
computer power is very unlikely to be ample for simulations with system sizes
larger than the correlation length in the foreseeable future.
Therefore, properly extracting the thermodynamic
limit from our data is essential. We will discuss the Monte Carlo simulation and derive
the theoretical finite-size scaling predictions in Sec.~\ref{sec2},
present results from our simulation and the data analysis in
Sec.~\ref{sec3}, and conclusions with discussions in Sec.~\ref{sec4}.

\section{Theoretical background}
\label{sec2}

\subsection{Monte Carlo simulation and traditional finite-size scaling}
At zero temperature, the system has a spin-flop transition at
$H_c=4J\sqrt{1-\Delta^2}$. The system is in the fully aligned AF
configuration for $H<H_c$ and is in the spin-flop configuration
for $H>H_c$. As $H \rightarrow H_c^+$, the spin-flop configuration
approaches $S_i^z =\sqrt{1-\Delta \over 1 + \Delta}$ and
$S_i^{x,y} = (-1)^in^{x,y}\sqrt{2\Delta \over 1+\Delta}$, where
$(n^x, n^y)$ is an arbitrary unit vector. Obviously, the spin-flop
transition is a first order phase transition at zero temperature,
since the order parameter (staggered magnetization) exhibits a
discontinuity. In scenario (a) in Fig.1, the spin-flop line starts
from $H_c$ at $T=0$, shows a small positive slope as the
temperature increases, and eventually develops into an umbilicus,
where the AF-paramagnetic phase boundary and the XY-paramagnetic
phase boundary are clearly separated.  For the other possibilities
in 2 dimensions, two phase boundaries might well be indistinguishable from
each other but yield the same ``effective'' finite temperature
behavior.

We perform Monte Carlo simulations that scan the magnetic field
across the spin-flop transition at constant temperatures. The
random spin configurations are generated with the heatbath
algorithm,~\cite{Miyatake86} which has been recently shown to be
the optimal single spin algorithm for Heisenberg
spins.~\cite{Loison04, Zhou04} It is a rejection-free algorithm
due to the fact that the probability distribution
$P(\bv{S})\propto \exp(\bv{h}\cdot\bv{S}/T)$ of a single spin
subject to a magnetic field can be produced with uniform random
number generators without rejecting trial flips. As in
Ref.~\onlinecite{Zhou04}, we simultaneously perform two
simulations with different initial configurations, the equilibrium
is achieved by letting these two simulations run until their order
parameters are almost equal. Each simulation then runs for $10^6$
to $10^7$ Monte Carlo steps per spin at the given temperature and
magnetic field, and is allowed to accumulate substantially more
data near the critical magnetic field. The data for energy,
magnetization, and staggered magnetization are stored for
histogram reweighing. The difference in observables measured from
independent simulations are used to estimate the error bar.
Totally we have spent about 1 CPU year and accumulated 27GB data,
which represents a modest computing load on a decent Linux
cluster.

We calculate the ensemble average of the staggered magnetization:
\begin{equation}
\bv{M}^\dagger = {1\over L^2} \sum_i (-1)^i \bv{S}_i,
\end{equation}
where the sign is different on two sublattices.
We define the tilt angle $\theta$ between
the $z$ axis and $\bv{M}^\dagger$ as
\begin{equation}
  \theta = \cos^{-1} {|M_z^\dagger|\over |M^\dagger|},
\end{equation}
where the value of $\theta$ is restricted to $[0, \pi/2]$ due to the
inversion symmetry of the staggered magnetization, and calculate
its probability distribution.
The Binder cumulant for an arbitrary observable $X$ is defined as
\begin{equation}
  U_4(X) = 1 - { \left<X^4\right>\over 3 \left<X^2\right>^2},
\end{equation}
where $\left<\cdots\right>$ denotes the ensemble average.
The susceptibility per spin of $M^\dagger_z$ is defined as
\begin{equation}
\chi = {L^2\over T }\left[\left<(M^\dagger_z)^2\right> -
\left<|M^\dagger_z|\right>^2\right],
\end{equation}
and the specific heat per spin is defined as
\begin{equation}
c_v = {L^2\over T^2 }\left(\left<E^2\right> -
\left<E\right>^2\right).
\end{equation}
where $E$ is the internal energy per site.

In order to determine the type of phase transitions, we performed
the traditional finite-size scaling analysis. In the case of an
Ising-like second order phase transition, the specific heat
exponent $\alpha = 0$, because $c_v$ is logarithmically divergent
near the critical temperature or critical field, and its maximum
value increases as $\ln L$. The localization length exponent $\nu
= 1$; the spontaneous magnetization exponent $\beta = 1/8$; and
the susceptibility exponent $\gamma = 7/4$.  In case of a first
order phase transition, these exponents can also be defined
properly, and they have a different set of values:~\cite{Binder84}
$\alpha =1, \beta = 0, \nu = 1/2, \gamma = 1$. (In general $\nu =
1/d$, where $d$ is the dimensionality of the system.) Near the
phase transition point, it is convenient to define the reduced
temperature $t = T/T_c-1$ and reduced field $h = H/H_c-1$.  Finite
size scaling theory predicts the following scaling relations:
\begin{eqnarray}
  \label{eq7}
  \left<(M^\dagger_z)^2\right>_L &=& L^{-2\beta/\nu} {\cal M} (tL^{1/\nu}),\\
  \label{eq8}
  \left<\chi \right>_L &=& L^{\gamma/\nu} {\cal C} (tL^{1/\nu}).
\end{eqnarray}
Similar relations hold for $(M^\dagger_x)^2+(M^\dagger_y)^2$ and $c_v$, as well as for
transitions with $h$ as the control variable. In particular, for $c_v$ of Ising-like
transition, the prefactor $L^{\alpha/\nu}$ is replaced by $C+\ln L$, where $C$ is a constant.

\subsection{Low temperature finite-size scaling predictions }
As we shall see from the results in the next section, neither the scaling relations
for the first order phase transition nor those for second order phase transitions fit
our data perfectly well. The first order phase transition apparently seems to be more
consistent though. To resolve these discrepancies, we return to the RG calculation
of bicritical phenomenon in $2+\epsilon$ dimensions. The long length scale physics is expected
to be governed by the non-linear $\sigma$ model with an anisotropy term:~\cite{Pelcovits76,Nelson77}
\begin{equation}
\label{eq9}
{\cal H}_\sigma = - {1\over 2T} \int d^{d}x \left[ (\partial_\mu \bv{\pi})^2+(\partial_\mu \sigma)^2 + g \sigma^2 \right],
\end{equation}
where the $O(3)$ spin field $(\pi_1(x),\pi_2(x),\sigma)$ satisfies
the constraint $\bv{\pi}^2+\sigma^2=1$ and the anisotropy constant
$g \propto H-H_c$. [Note: we follow the convention in
Ref.~\onlinecite{Nelson77} here, so the Boltzmann factor is
$\exp({\cal H}_\sigma)$.] In general, $g$ is expected to be a
linear combination of both magnetic field and temperature
contributions:~\cite{Fisher75} $g = a(H^2-H_c^2) + b (T-T_c)$.
The spin-flop line (defined by $g=0$) of our model is almost
horizontal, with a very small negative $b$. In the scenario shown
in Fig.~\ref{fig0}(b), $H_c$ actually falls in the paramagnetic
phase between two phase boundaries. The RG calculation can proceed
with either the momentum shell renormalization
technique~\cite{Nelson77} or Polyakov's
approach.~\cite{Polyakov75, Pelcovits76} Same results can also be
derived with Br\'ezin and Zinn-Justin's method.~\cite{Brezin76,
Zinn-Justin76} The RG flow differential equations for renormalized
temperature $T(l)$ and anisotropy constant $g(l)$  are
\begin{eqnarray}
\label{eq10}
  {dT(l)\over dl} &=& -\epsilon T(l) + {T(l)^2\over 2\pi}\left[ n-3 + {1\over 1+g(l)} \right],\\
\label{eq11}
  {dg(l)\over dl} &=& 2g(l) - {T(l) \over \pi} { g(l) \over 1+g(l)},
\end{eqnarray}
where $\epsilon = d -2$. For our present purpose, it is sufficient
to set $\epsilon = 0, n = 3$ and keep the lowest order terms of
$g$ in these  equations. The approximate solution then reads
\begin{eqnarray}
  \label{eq12}
  T(l) &=& { 2\pi T\over 2\pi -Tl}, \\
  \label{eq13}
  g(l) &=& g e^{2l} T^2/T(l)^2 = g e^{2l} [1-Tl/(2\pi)]^2.
\end{eqnarray}
Additionally, the equation for the renormalization constants for
$\bv{\pi}$ and $\sigma$ can be derived from
Ref.~\onlinecite{Nelson77}:
\begin{eqnarray}
 { d\ln \zeta_\pi \over dl }&=& d - {T(l)\over 4\pi} { g(l)+2\over g(l)+1},  \\
 { d\ln \zeta_\sigma \over dl } &=& d - {T(l)\over 2\pi}.
\end{eqnarray}
Using Eq.~(\ref{eq12}) and $d=2$, for negligible $g$, one finds
that $\zeta_\bv{\pi}$ and $\zeta_\sigma$ are both given by
\begin{equation}
  \label{eq16}
  \zeta_\sigma = \zeta_\pi = e^{2l}\left(1-{T l \over 2\pi}\right).
\end{equation}
These results imply that for $g>0$, the renormalization flow goes
to $g=+\infty$, which corresponds to the XY model; for $g<0$, it
goes to $g=-\infty$, which corresponds to the two-dimensional
Ising model; for $g=0$, it goes to the high temperature
paramagnetic phase, which is consistent with the Mermin-Wagner
theorem.~\cite{Mermin66} Conventionally, the thermodynamic
quantities are calculated by integrating the renormalization flow
equation to a characteristic scale $l^*$, for which $g(l^*)=
O(1)$.  The corresponding thermodynamic quantities can be
calculated by other means, e.g. perturbation theory. Then a
trajectory integral ``matching''
formalism~\cite{Nelson75,Nelson76,Rudnick76,Nelson77} is used to
obtain these quantities at the original scale $l=0$.  For example, 
the susceptibility has the scaling form $\chi(T,g) \sim e^{4\pi/T}\Phi(ge^{4\pi/T})$ for
small $T$ and $g$ in the thermodynamic limit,~\cite{Nelson77}
which is analogous to the crossover phenomenon
of a finite temperature bicritical point.\cite{Nelson74, Fisher74} 
The Ising or XY critical behavior is contained in the scaling function $\Phi$,
which is calculated from the effective Ising or XY model (depending on the sign of $g$) 
at the scale $l^*\sim e^{4\pi/T}$.
Due to the finite size of the systems in our simulations, the above RG flow
cannot go very far in the RG flow diagram for an infinite system,
but only to the scale set by the system size $l' \approx \ln
(L/a)$, where $a$ is the microscopic lattice constant (fixed to be unity). For our
model, $L=100a$ is not a large system size compared to the
correlation length, so we expect the lowest order solutions
Eqs.~(\ref{eq12}) and (\ref{eq13}) are sufficient. In the real
space RG terminology,~\cite{Kadanoff76} at scale $l'$, the spins
in a block of linear size $e^{l'}$ are effectively replaced by a
single block spin, therefore, we expect our simulation with system
size $L$ in a magnetic field sufficiently close to the critical
field is governed by the simple Hamiltonian
\begin{equation}
\label{eq17}
  {\cal H}_{\rm eff} = - q \sigma^2,
\end{equation}
where the coefficient $q$ is given by
\begin{equation}
\label{eq18}
  q ={ g(\ln L) \over 2 T(\ln L)} = {gL^2\over 2T} \left(1-{T\ln L \over 2\pi}\right)^3.
\end{equation}
Near the spin-flop transition, $g = k(T,L)h + O(h^2)$, where $h$
is the reduced field, and we expect the coefficient $k(T,L)$ to
weakly depend on $T$ and $L$. As a result, the free energy should
appear as a function of $hT^{-1}L^2[1-T\ln L /(2\pi)]^3$. One can
define a ``correlation length exponent'' $\nu = 1/2$, based on the
leading $L^2$ dependence. The staggered magnetization also has to
contain a factor from the spin renormalization constant:
\begin{eqnarray}
\label{eq19}
  &&\left<(M^\dagger_z)^2\right>_L = L^{-4} \zeta_\sigma^2\left<\sigma^2\right>_{{\cal H}_{\rm eff}},\\
\label{eq20}
  &&\left<(M^\dagger_{xy})^2 \right>_L = \left<(M^\dagger_x)^2 +(M^\dagger_y)^2 \right>_L =L^{-4} \zeta_{\bv{\pi}}^2\left<\pi^2\right>_{{\cal H}_{\rm eff}},
\end{eqnarray}
where $\left<\dots\right>_{{\cal H}_{\rm eff}}$ denotes the thermal average with respect to the
effective Hamiltonian Eq.~(\ref{eq17}).
When $\zeta_\sigma/\zeta_\pi \approx 1$, both of them
have the scaling form:
\begin{equation}
\label{eq21}
  \left(1-{T\ln L \over 2\pi}\right)^2
 {\cal F}_{\parallel, \perp}\left[{hL^2\over T}\left(1- {T\ln L \over 2\pi}\right)^3\right].
\end{equation}
Furthermore, the distribution of $\bv{M}^\dagger$ should be on a sphere of radius
$1-T\ln L / (2\pi)$; the distribution of its tilt angle has the form
\begin{equation}
\label{eq22}
  P(\theta) = Y^{-1} \exp(-q \cos^2\theta)\sin\theta,
\end{equation}
where $Y$ is a normalization constant. At the critical field $g =
h = 0$, the above distribution is uniform. Consequently the
critical Binder cumulant for $\cos\theta$ is $U_4(\cos\theta) =
2/5$. $U_4(M^\dagger_z)$ should also be close to this value, with
a  small negative correction due to the longitudinal fluctuation
of $\bv{M}^\dagger$. The maximum susceptibility corresponds to the
maximum fluctuation in $\sigma$, so the following scaling formula
is expected:
\begin{equation}
\label{eq23} \chi = {\zeta_\sigma^2\over L^2} {\cal
X}\left[{hL^2\over T}\left(1-{T\ln L \over 2\pi}\right)^3\right].
\end{equation}
One can again define a susceptibility exponent $\gamma = 1$ based
on the leading $L^2$ dependence, which is assumed to be
$L^{\gamma/\nu}$. The critical exponent ratio is then identical to
that for first order phase transition; therefore, the logarithmic
corrections are required to determine whether the spin-flop
transition is a first order phase transition or not.

To calculate the specific heat, the standard approach is to derive it from
the free energy $ f = - L^{-2}\ln Z$. As in Ref.~\onlinecite{Nelson77},
we perform a trajectory integral from $l=0$ to $l=l^*$
to evaluate the free energy. For the clarity of our discussion, we
reproduce the trajectory integral formula from Ref.~\onlinecite{Nelson77} here:
\begin{equation}
\label{eq24}
f(T,g) = \int_0^{l^*} e^{-dl}G_0(l)dl + e^{-dl^*}f[T(l^*),g(l^*)],
\end{equation}
in which we set $d=2$ in our calculation and the kernel $G_0$ depends on both the
on-shell Green's function and the spin renormalization factors.
With the final scale $l^* = \ln L$ and the approximate solution Eqs.~(\ref{eq12})
and (\ref{eq13}), this trajectory integral can be evaluated analytically
with the technique in Appendix A of Ref.~\onlinecite{Rudnick76}.
Among many terms in the result, we are particularly interested in those of the form
$L^{-2}f_s[T(l^*),g(l^*)]$, for reasons which will soon become clear.
It turns out that the only such term is $f_s[g(l^*)] = -\ln[1+g(l^*)]$.
The last term in Eq.~(\ref{eq24})  is given by the single spin Hamiltonian Eq.~(\ref{eq17}):
\begin{equation}
\label{eq25}
  L^{-2} F\left[ {gL^2\over 2T}\left(1-{T \ln L\over 2\pi}\right)^3\right],
\end{equation}
where
\begin{equation}
  F(q) = -  \ln  \int_0^\pi  \exp(-q \cos^2\theta) 2\pi \sin\theta d\theta.
\end{equation}
The divergent term in the specific heat is given by
\begin{equation}
\label{eq27}
  c_v = -{\partial \over \partial T}\left[T^2 L^{-2}{\partial (F+f_s) \over \partial T }\right].
\end{equation}
One would expect the leading divergent term in Eq.~(\ref{eq27}) is
proportional to $L^2g^2$, with multiplicative  logarithmic
corrections. However, obviously on the spin-flop line $g=0$, this
term  vanishes. This dilemma is solved by noting that $T$ in
Eq.~(\ref{eq27}) is the real temperature at which the simulation
is performed, while $T$ in Eqs.~(\ref{eq24}) and (\ref{eq25}) is a
renormalized temperature for the long length scale effective
Hamiltonian, and from now on it will be denoted $T^*$. The
spin-flop line does not follow a constant temperature, in other
words, the effective anisotropy $g$ is also a function of $T$. If
we change $T$ in the simulation, we actually change the effective
anisotropy $g$, which could drive the system to cross the
spin-flop line. Therefore, the partial derivative in
Eq.~(\ref{eq27}) should be written as
\begin{equation}
{\partial \over \partial T} = a {\partial \over \partial T^*}+ b {\partial \over \partial g}.
\end{equation}
Two derivatives $\partial/\partial g$ acting on $L^{-2}F$ would produce a divergent
term proportional to $L^2 [1-T^*\ln L /(2\pi)]^6$. Similarly,
$f_s$ results in a divergent term proportional to $L^2 [1-T^*\ln L /(2\pi)]^4$.
After all these considerations, the divergent part of the specific heat is expected to have the
following scaling form:
\begin{equation}
\label{eq29}
  c_v = L^2 x^6 {\cal C}_6\left(hL^2x^3\right) +
   L^2 x^4 {\cal C}_4\left( hL^2 x^3\right),
\end{equation}
where $x = 1-T^*\ln L /(2\pi)$. If the spin-flop line is perfectly horizontal,
i.e. $b=0$, there will not be a peak in the specific heat. We expect the ${\cal C}_6$
term to be dominant in Eq.~(\ref{eq29}),
because $F$ also contributes a factor $T^{-2}$ after being differentiated
twice, on the other hand, a straightforward calculation shows that the ${\cal C}_4$
term does not show a peak at $g=0$, but an asymmetric background.

The above trajectory integral actually connects the intermediate renormalized
Hamiltonian Eq.~(\ref{eq9}) to the final renormalized Hamiltonian Eq.~(\ref{eq17}).
In doing so, we have actually ignored a similar trajectory integral which connects
the bare Hamiltonian Eq.~(\ref{eq1}) to the intermediate Hamiltonian Eq.~(\ref{eq9}).
Although we cannot write down an analytic expression for its integrand and its
final contribution to the free energy, we expect it to depend on the lattice constant,
which is the smallest length scale in the system, and an intermediate length scale
at which Eq.~(\ref{eq9}) is valid, but not on the largest length scale $L$.
Therefore, it cannot possibly give rise to an $L^2$ divergence. For the divergent
terms in $c_v$, it is hence justified to ignore this precursory
trajectory integral.

Unlike the conventional finite-size scaling, the above finite-size
scaling relations are not expected to hold for very large system
sizes, because for sufficiently large $L$, the phase boundaries of
the AF and XY phases become discernible. This obviously happens
when $T\ln L \sim 2\pi$, i.e. the spin renormalization constants
become very small, corresponding to a disordered phase. For either
large $l$ or large $g$, the higher order terms in Eqs.~(\ref{eq10})
and (\ref{eq11}) become non-negligible; therefore, the approximate
solution Eqs.~(\ref{eq12}) and (\ref{eq13}) and the above finite-size 
scaling relations are no longer sufficient. Thus, we expect
the above finite-size scaling analysis are valid for small
effective anisotropy, i.e. $H\approx H_c$, and system sizes
smaller than the correlation length.

\section{Data and analysis}
\label{sec3}
\subsection{Raw data and the apparent first order  phase transition}
\label{sec3a}

\begin{figure}[t]
\includegraphics[width = 1.0\columnwidth]{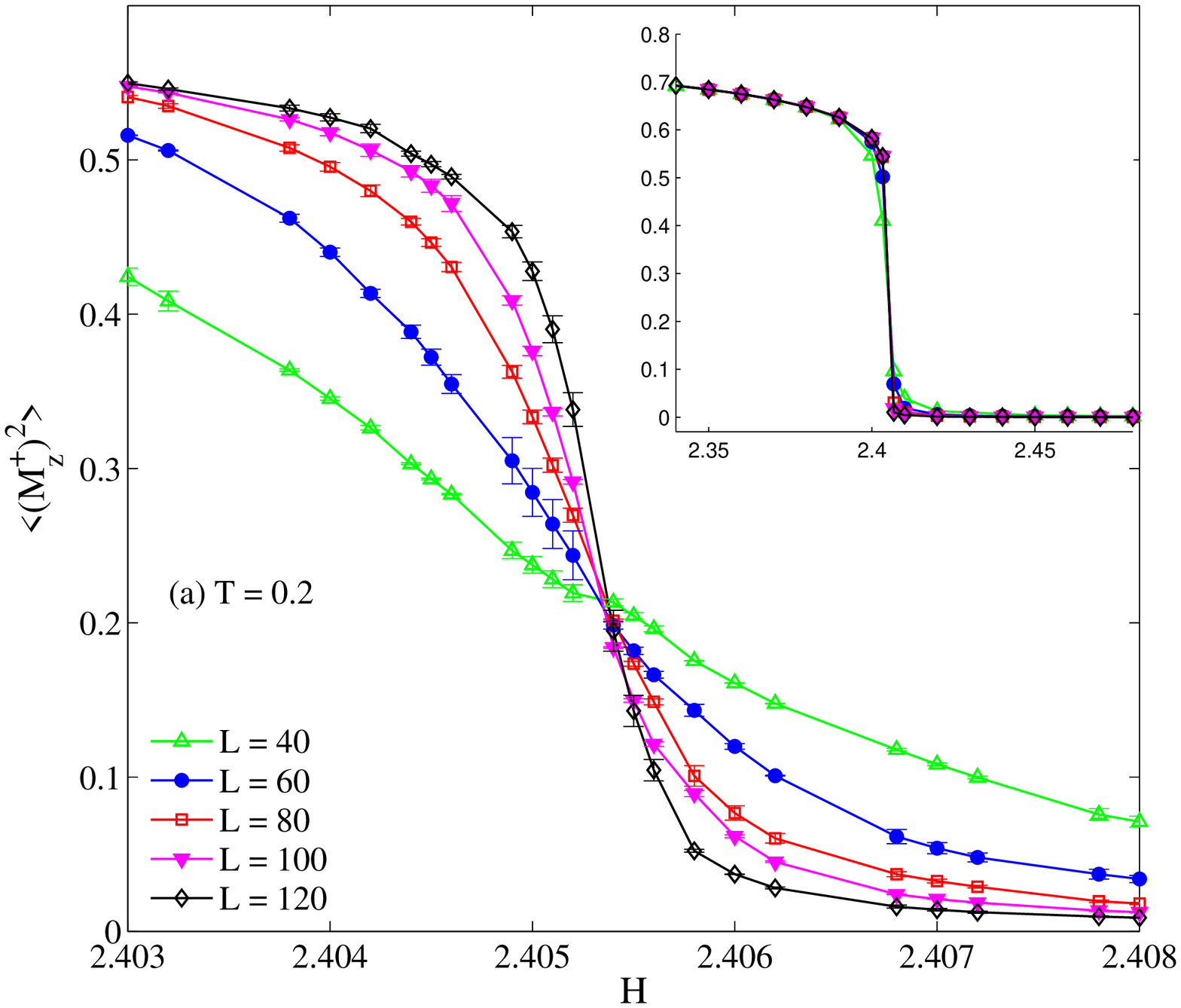}
\includegraphics[width = 1.0\columnwidth]{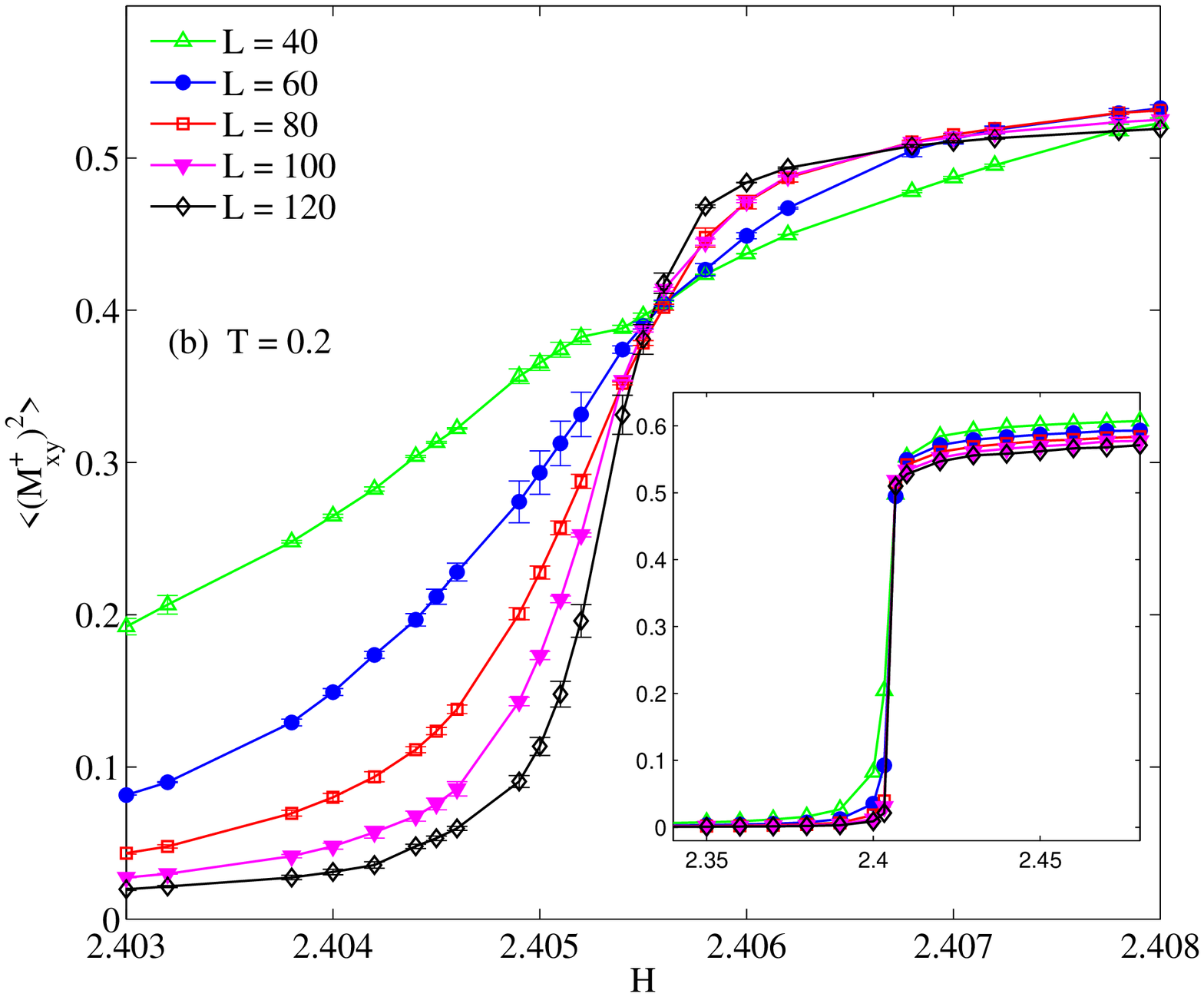}
\caption{(color online) Staggered magnetization across the finite
lattice spin-flop line at $T=0.2$. Each group of nearby data
points are from one simulation performed with single fixed
magnetic field. Histogram reweighing was used to calculate
observables in slightly different magnetic fields. The insets show
the data measured from a larger range of magnetic field.}
\label{fig1}
\end{figure}

We first present some representative raw data for the staggered
magnetization from simulations with different system sizes, and we
will then show traditional finite-size scaling plots.
Figure~\ref{fig1} shows the staggered magnetization at $T=0.2$ for
magnetic fields on both side of the finite lattice spin-flop
transition. The spin-flop transition can be clearly observed, and
when viewed at moderate resolution (see the insets) the data
suggest a 1st-order transition. Both
$\left<(M^\dagger_z)^2\right>$ and
$\left<(M^\dagger_{xy})^2\right>$ reach either zero or a large
value quickly as $H$ deviates from $H_c$, which is why we have to
zoom in on a very narrow range (0.005 in width) of the magnetic
field to see the finite-size effect for relatively large systems.
The critical magnetic field that separates the XY phase from the
AF phase can already be estimated rather accurately from the size
dependence seen in Fig.~\ref{fig1}. There is no noticeable sign of
an intermediate phase which has zero expectation values for both
order parameters.  There is also no sign of a region with nonzero
values for both order parameters. In the XY phase, we see from the
inset of Fig.~\ref{fig1}(b) that the saturated values of the order
parameter are smaller for larger systems, which is consistent with
the property of the XY phase. (We expect the saturated value to
decrease as $L^{-2\eta}$, where $\eta$ is a temperature dependent
exponent.~\cite{Thouless73,Peczak91}) On the other side, in the AF
phase, the inset of Fig.~\ref{fig1}(a) shows that
$\left<(M^\dagger_z)^2\right>$ calculated with different system
sizes quickly converges to the nonzero thermodynamic limit.

\begin{figure}[t]
\includegraphics[width = 1.0\columnwidth]{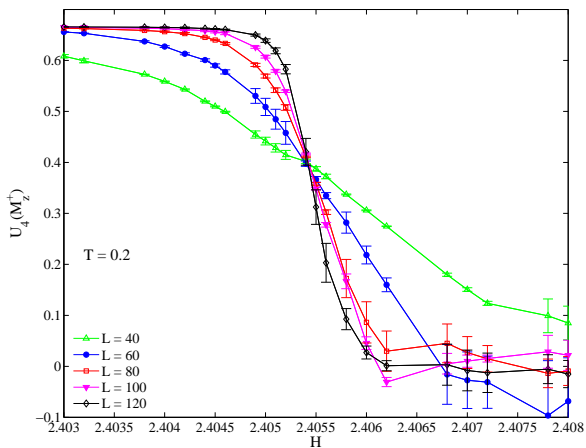}
\caption{(color online) Binder cumulant for various sizes at $T=0.2$,
the crossing points of these curves are near $H=2.4055$, $U_4$ at the
crossing point is different from the universal value for Ising-like
transition.}
\label{fig2}
\end{figure}

If the spin-flop transition contains an Ising-like transition,
then we would expect to see the universal value of the Binder
cumulant $U_4^* \approx 0.618$ at the critical magnetic field,
therefore we plot $U_4(M^\dagger_z)$ in Fig.~\ref{fig2} to see if
it indicates an Ising-like second order phase transition. However,
the curves in Fig.~\ref{fig2} cross each other at values close to
0.4, clearly below the Ising universal value. There is not any
systematic trend indicating that the crossing point moves up with
increasing system size. This result is in agreement with
Ref.~\onlinecite{Leidl05}, where it is shown that only when
$T>0.4$ do the crossing points in the Binder cumulant curves start
to move towards the Ising universal value. As
Ref.~\onlinecite{Leidl05} has shown in the phase diagram, the
AF-paramagnetic boundary and XY-paramagnetic boundary are clearly
separated for $T>0.4$. We have also calculated the Binder cumulant
for $T=0.1, 0.265$, and 0.33, and seen results similar to
Fig.~\ref{fig2}. So it is indeed tempting to consider the
possibility of a bicritical point which exists between $T=0.33$
and $T=0.4$.

\begin{figure}[th]
\includegraphics[angle=270,width=1.0\columnwidth]{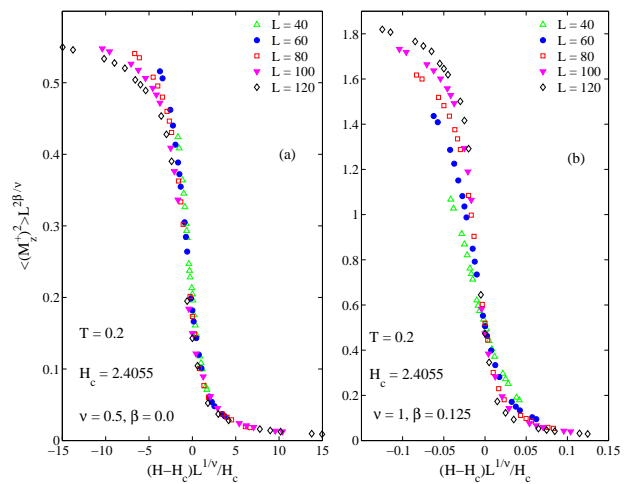}
\caption{(color online) Finite-size scaling of $M^\dagger_z$. (a) Exponents
used belong to first order phase transition; (b) Ising exponents are used.
Error bars are not plotted since most of them are smaller or equal to the
size of the symbols, as shown in Fig.~\ref{fig1}.}
\label{fig3}
\end{figure}

To see whether our data really fit the behavior expected for a
first order phase transition, we show the finite-size scaling
analysis of $M^\dagger_z$ in Fig.~\ref{fig3}. According to
Eq.~(\ref{eq7}), all the data points should collapse onto a single
curve if we choose the right exponents $\nu$ and $\beta$. The
field dependence of the data in both Fig.~\ref{fig1} and
Fig.~\ref{fig2} indicate that $H_c$ is between 2.405 and 2.406. In
Fig.~\ref{fig3}, we have fine tuned $H_c$ to collapse the data as
much as possible onto a single curve, and found that it is
impossible to do so with Ising exponents. With first order
exponents, the scaling plot Fig~\ref{fig3}(a) is clearly better.
However, systematic deviations are discernible in the low field AF
phase. In fact, if we allow $\beta$ to have a small value, we have
found that the scaling plot looks best with $\beta=0.031$. The
same phenomenon has also been observed in traditional finite-size
scaling plots for $\left<(M^\dagger_{xy})^2\right>$.

Similar scaling analyses have been done for $T=0.1$, and 0.265. In
all those cases, the Ising exponents fail, but the first order
exponents work reasonably well, especially for $T=0.1$. If we
accept a small $\beta$ which goes to zero as $T$ goes to zero,
then at finite temperatures, the spontaneous staggered
magnetization does decay as $M^\dagger_z \propto |h|^\beta$ in the
thermodynamic limit ($L\rightarrow \infty$). We would indeed have
a second order phase transition. However the correlation length
exponent $\nu = 1/2$ is not consistent with this scenario. For
susceptibility and specific heat, finite-size scaling analysis has also
ruled out Ising-like transition as a possible scenario. Especially
for specific heat, its maximum value scales roughly as $L^2$ at
$T=0.1, 0.2$, which is glaringly different from the Ising-like
logarithmic behavior.

\subsection{Proper finite-size effect and scaling}

\begin{figure}[htbp]
\includegraphics[width=1.0\columnwidth]{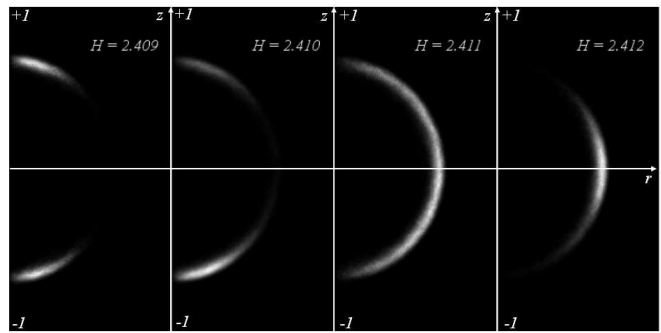}
\caption{ Probability distribution of the staggered magnetization across the spin-flop transition. $P(r,z)$ is proportional to the grey scale. The data are from simulations with $L=100$ at $T=0.265$.}
\label{fig4}
\end{figure}

To characterize the spin-flop transition, we first study the
probability distribution of the staggered magnetization to get a
vivid picture of the spin-flop transition. Due to the obvious
rotational symmetry around the $z$ axis, we plot in
Fig.~\ref{fig4} the probability distribution $P(r,z)$ of systems
with $L=100$ in cylindrical coordinates at $T=0.265$. It can be
clearly seen from this figure that $\bv{M}^\dagger$ is distributed
on a sphere of about 0.65 in radius. In the AF phase, the
distribution is confined to the north and south poles; in the XY
phase, the distribution is near the equator. One would naturally
define a {\it finite-size } critical field $H_c$ at which the
distribution is uniform on this sphere, i.e. zero effective
anisotropy. For $L=100$ and $T=0.265$, we have found $H_c \approx
2.411$ which is in agreement with the observation of
$\left<(M^\dagger_z)^2\right>$ and $U_4(M^\dagger_z)$. At this
critical field, $U_4(M^\dagger_z) \approx 2/5$, which can be
immediately calculated given this uniform distribution. The same
pattern of transition have been observed at different temperatures
with different system sizes. For larger systems and higher
temperatures, we have observed smaller spheres. Clearly, a finite-size 
effect can be extracted from these spherical distributions.

\begin{figure}[t]
\includegraphics[angle=270,width=0.9\columnwidth]{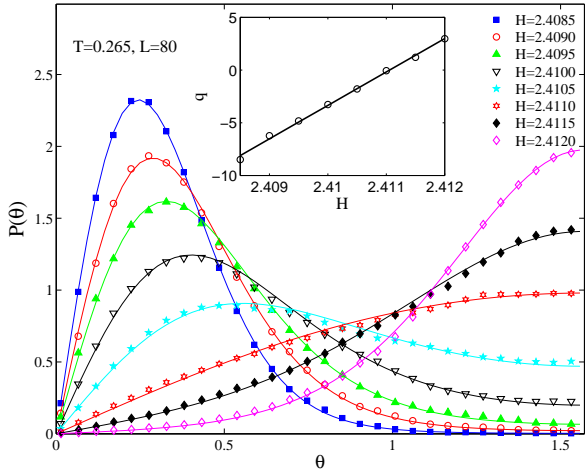}
\caption{(color online) Probability distribution of the tilt angle $\theta$. Symbols are calculated from the histogram of $\theta$ in the simulations, and the solid lines are curve fitting with Eq.~(\ref{eq22}). The fitting parameter $q$ is plotted in the inset, where the straight line is a linear fit.  The simulations were performed with $L=80$ at $T=0.265$.}
\label{fig5}
\end{figure}

To be more quantitative, we calculate the angular distribution of the staggered magnetization.
Figure~\ref{fig5} shows the
distribution of the tilt angle $\theta$ calculated from simulations at $T=0.265$ for $L=80$.
The data are fitted with Eq.~(\ref{eq22}), which only has one fitting parameter $q$.
Obviously this simple functional form fits the simulations very well.
The inset in Fig.~\ref{fig5} shows the fitting parameter, i.e. Eq.~(\ref{eq18}),
is a linear function of $H$, which is consistent with our assumption that $g\propto h$.
Thus, we have shown that close to $H_c$, the thermodynamics of the staggered magnetization
is indeed governed by the simple Hamiltonian Eq.~(\ref{eq17}).

The staggered magnetization clearly has a finite value at $H_c$ in
the simulations. If this is also true for the thermodynamic limit,
then we will have to choose between the scenario of a first order
phase transition, or that of an intermediate ordered phase with a
tetracritical point. However, this is not the case for
two-dimensional systems. Suppose we enlarge the simulation to a
size $nL$, since we have shown that an $L\times L$ block indeed
behaves as an effective Heisenberg spin in an anisotropic
potential, the long length scale physics should be captured by a
Hamiltonian of these block spins. Although microscopically, the
anisotropy may have different forms, at a large enough length
scale the $g\sigma^2$ term turns out to be the only dominant term,
as shown by the probability distribution in Fig.~\ref{fig5}. In
terms of RG terminology, it is the only relevant perturbation that
keeps the $z$ axis rotational symmetry. Since a similar anisotropy
in the kinetic energy, $g_1(\partial_\mu \sigma)^2$ has been found
to be irrelevant,~\cite{Pelcovits76} our simulations have indeed
justified Eq.~(\ref{eq9}) as an appropriate Hamiltonian to analyze
the spin-flop transition. Following the RG analysis outlined in
Sec.~\ref{sec2}, one would conclude that the behavior in the
thermodynamic limit is two second order phase boundaries and a
bicritical point at zero temperature, i.e. that which is shown in
Fig.~\ref{fig0}(b).

\begin{figure}[t]
\includegraphics[angle = 270,width=\columnwidth]{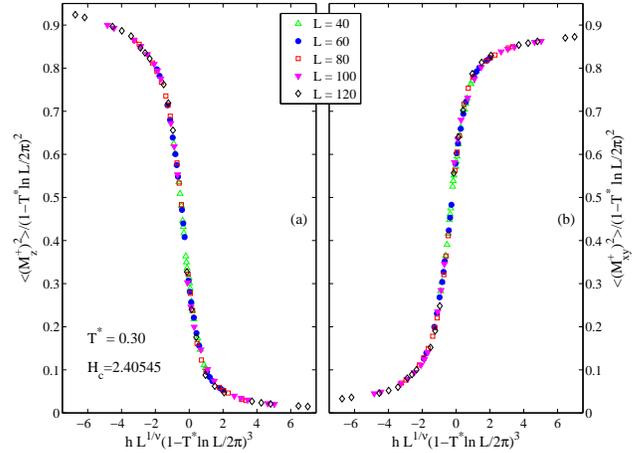}
\caption{(color online) Finite-size scaling analysis corresponding to Eq.~(\ref{eq21}) at $T=0.2$. (a) The AF order parameter; (b) the XY order parameter. We introduce $T^*$ as a fitting parameter which is slightly above the actual temperature at which the simulation is done. Error bars are smaller or equal to the size of the symbols.}
\label{fig6}
\end{figure}

\begin{figure}[b]
\includegraphics[angle = 270,width=1.0\columnwidth]{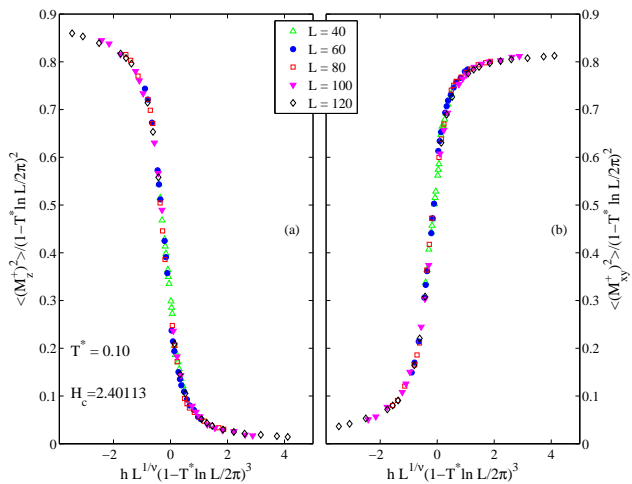}
\caption{(color online) The same as Fig.~\ref{fig6}, but for simulations at $T=0.1$. Unlike Fig.~\ref{fig6}, we have used $T^* =T$ here.}
\label{fig7}
\end{figure}

Now we perform a finite-size scaling analysis in order to examine
whether or not the logarithmic corrections in Eq.~(\ref{eq21}) can
be seen in the data from the simulations. The renormalized
temperature $T$ in the effective Hamiltonian Eq.~(\ref{eq9}) is,
in general, different from the temperature at which the simulation
is performed. Therefore, we are allowed to use an extra fitting
parameter $T^*$ in Fig.~\ref{fig6} to make the data points
collapse better on a single curve. We find that all the data
points do collapse onto a single curve in both Fig.~\ref{fig6}(a)
and \ref{fig6}(b) with the choice $T^*=0.3$. Actually, if this
renormalized temperature is not introduced, Fig.~\ref{fig6} still
looks clearly better than Fig.~\ref{fig3}(a), where we assume a
first order phase transition.
In the transition region, we also expect
$\left<(M^\dagger_z)^2 + (M^\dagger_{xy})^2\right>$ to be nearly a constant as seen in Fig.~\ref{fig4}.
This is also consistent with Fig.~\ref{fig6} where the decrease in
$\left<(M^\dagger_z)^2\right>$ is compensated by the increase in $\left<(M^\dagger_{xy})^2\right>$.
 A small $\beta$ improves Fig.~\ref{fig3}(a) because a
logarithmic correction enters by the Taylor expansion for small
$\beta$: $L^{2\beta/\nu} \approx 1 + 2\beta \nu^{-1}\ln L$. Error
bars are not plotted in Fig.~\ref{fig6} and the other scaling
plots, because most of them are smaller or equal to the size of
the symbols used in these figures. They can be found in
Fig.~\ref{fig1} and \ref{fig2}. Figure~\ref{fig7} shows the same
scaling analysis for simulations performed at $T=0.1$. We have
found that this scaling plot is very clear even without
introducing a renormalized temperature. The critical field $H_c$
is fixed up to six significant digits in Fig.~\ref{fig6} and
\ref{fig7} by closely examining the central critical region of
very small $h$. However, one should not push this too far because
the apparent existence of a single critical field is only a finite-size 
effect. For large systems, there are two separate critical
fields for the AF phase boundary and the XY phase boundary
respectively. The same finite-size scaling analysis at $T=0.265$
still produces a very good single curve; however, the best scaling
is achieved by allowing $H_c$ for the XY phase to be slightly
above that for the AF phase, behavior which could be understood as
a sign of separating phase boundaries.

\begin{figure}[t]
\includegraphics[angle = 270,width=1.0\columnwidth]{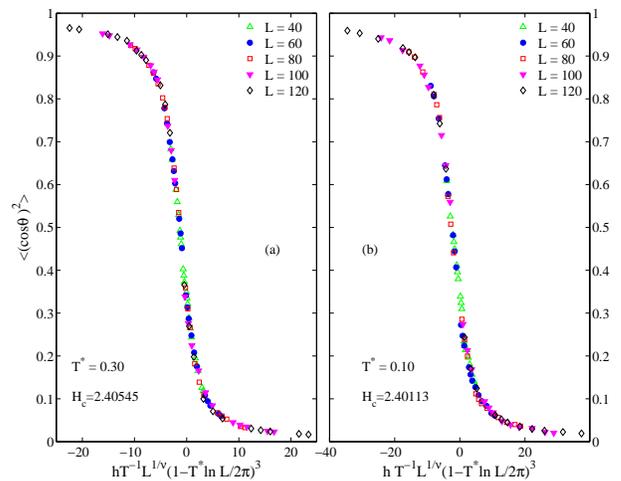}
\caption{(color online) Finite-size scaling plot of the effective $\sigma$ spin component, (a) from simulations at $T=0.2$, (b) from simulations at $T=0.1$. Instead of adjusting $T^*$, we have simply set $T^* = T$. (a) and (b) are expected to displays the same curve, see text for details. Error bars are smaller or equal to the size of the symbols.}
\label{fig8}
\end{figure}

The logarithmic correction for $\bv{M}^\dagger$ comes from the spin renormalization constants in
Eqs.~(\ref{eq19}) and (\ref{eq20}). The effective spin $\sigma$ and $\bv{\pi}$ can be calculated
directly from $\bv{M}^\dagger$, given that $\zeta_\bv{\pi} \approx \zeta_\sigma$. One would expect
that the finite-size scaling of $\left<\sigma^2\right>$ only involves a scaling of the magnetic field. In fact, Eqs.~(\ref{eq17}) and (\ref{eq18})
imply that $\left<\sigma^2\right>_L$ for different sizes as well as
different temperatures can collapse on to a single curve, provided that $k(T,L) = h/g$ depends
weakly on $T$ and $L$. Figure~\ref{fig8} plots
$\left<\sigma^2\right>_L$ versus $hT^{-1}L^2(1-T\ln L/2\pi)^3$, and shows
that this is indeed true in the simulations. We have also calculated the Binder cumulant
$U_4(\cos\theta)$. Since the longitudinal fluctuation in $\bv{M}^\dagger$ is
completely projected out, $U_4(\cos\theta)$ is exactly 2/5 at the crossing point, which
corresponds to  the critical magnetic field for finite-size systems.

\begin{figure}[t]
\includegraphics[angle = 270,width=1.0\columnwidth]{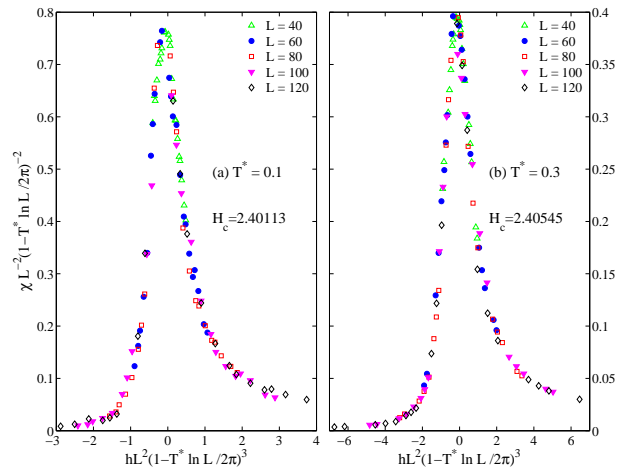}
\caption{(color online) Scaling plot for susceptibility near the spin-flop transition according to Eq.~(\ref{eq23}). (a) at $T=0.1$, (b) at $T=0.2$.}
\label{fig9}
\end{figure}

As for the susceptibility, we test Eq.~(\ref{eq23}) by plotting
$\chi L^{-2}(1-T\ln L /2\pi)^{-2}$ versus $hL^2(1-T\ln L/2\pi)^3$ in
Fig.~\ref{fig9}. For $L=40, 60$, and 80, our data collapse onto
single curves at $T=0.1$ and $T=0.2$ very well. For two larger
sizes, $L=100$ and $L=120$, although we have fewer data points
near the peak of the susceptibility and a few data points have
larger statistical errors than others, they appear to fall on the
same curve reasonably well. We have also plotted the data assuming
finite-size scaling for an Ising-like transition or a first order
phase transition but found none of those works better than
Fig.~\ref{fig9}.

\begin{figure}[t]
\includegraphics[angle = 270,width=1.0\columnwidth]{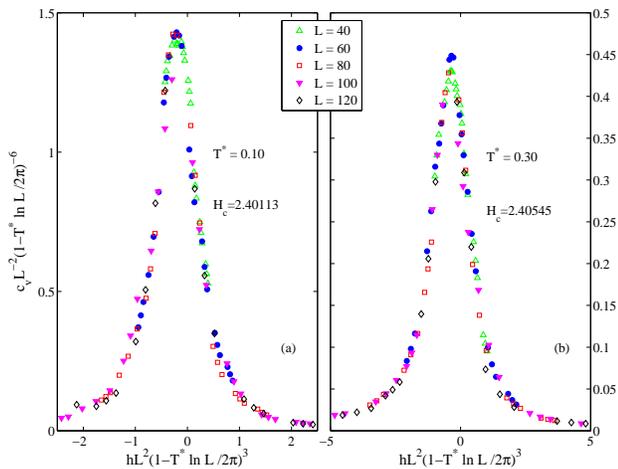}
\caption{(color online) Scaling plot for specific heat near the spin-flop transition according to Eq.~(\ref{eq29}). (a) at $T=0.1$, (b) at $T=0.2$. }
\label{fig10}
\end{figure}

Finally, in Fig.~\ref{fig10} we show the finite-size scaling plots
for specific heat which follow Eq.~(\ref{eq29}). Only the first
${\cal C}_6$ term has been considered here. Most of the data
points collapse quite well onto a single curve in both
Fig.~\ref{fig10}(a) and \ref{fig10}(b), and this verifies the
validity of our finite-size scaling analysis. We have made the
same scaling plot for specific heat at $T=0.265$ and 0.33, and
found none of them obeys this scaling formula. This is obviously
because at higher temperatures, phase boundaries of the XY and AF
phases start to separate. In fact, those data at higher
temperatures are consistent with an Ising transition. As in other
figures, a group of nearby data points are calculated from the
same simulation with histogram reweighing. The disadvantage of
histogram reweighing is that the statistical error in one
simulation is replicated by several data points. Therefore, the
deviations observed in Fig.~\ref{fig10} as well as Fig.~\ref{fig9}
are not real ``systematic'' deviations, but an artifact of
histogram reweighing.

\section{Discussion}
\label{sec4}

Although near the critical magnetic field, the simulation is
hindered by the huge correlation length, so that two separate
second order phase boundaries may never be revealed, the staggered
magnetization behaves as a rigid spin of nearly fixed length
subject to an anisotropic potential. The length of the spin, and
the form and strength of the anisotropy are predictions of the RG
calculation, which have been confirmed by our Monte Carlo
simulations. Logarithmic corrections of particular forms, which
are absent in either first order phase transition or second order
phase transitions, have been found in the Monte Carlo simulations
in complete agreement with our theoretical predictions. For the
phase diagram of the two-dimensional XXZ Heisenberg
antiferromagnet with an easy axis, we have thus reached the
conclusion that in the thermodynamic limit, the bicritical point
appears at $T=0$, with an exponentially narrow paramagnetic phase
sandwiched between the low field AF phase and the high field XY
phase.  The location of the bicritical point is ``hidden'' from
detection by ``ordinary'' means because of finite-size effects
that must be carefully analyzed.  In the simulation of any finite-size 
system, the staggered magnetization remains non-zero on the
``spin-flop line'' which only becomes two separate phase
boundaries in the thermodynamic limit. The Ising-like or XY-like
critical behavior near the spin-flop transition is also expected
to be seen in an exponentially narrow range of magnetic field. Our
results should be valid for all two-dimensional Heisenberg
antiferromagnet or ferromagnet with short range interaction and
easy axis anisotropy of different forms, since their long length
scale physics is described by the same Hamiltonian. Due to the
finite experimental resolution, including effects from the
inhomogeneity in magnetic field and disorder in samples, and
possible crossover to three dimensional behaviors, we expect
experiments would only reveal apparent first order spin-flop
transitions and an apparent bicritical point at finite
temperature. Also symmetry breaking perturbations, e.g. crystal
fields of square symmetry, are highly relevant to the bicritical
phenomena in $2+\epsilon$ dimensions as well as the intricate
correlations in the XY phase.~\cite{Pelcovits76,Rastelli04} As a
result, both the AF phase and the ``XY'' phase might have discrete
symmetries, so that the two second order phase boundaries are
likely to be reduced to a first order phase boundary.

\section{Conclusion}
\label{sec5}

We have carried out extensive Monte Carlo simulations for a two-dimensional 
anisotropic Heisenberg antiferromagnet near the
spin-flop transition and performed a finite-size scaling analysis
based on the anisotropic non-linear $\sigma$ model in the regime
where the correlation length is much larger than the system size.
 We have found that, although the finite-size effects tend to obscure
 the asymptotic behavior, the results of the Monte Carlo simulations are perfectly
consistent with the RG calculation for the spin-flop transition
and that the bicritical point is at $T=0$.

\begin{acknowledgments}
We thank K. Binder and S. Mitchell for very helpful discussions and suggestions.
This research is supported by the Department of Energy through the Laboratory Technology
Research Program of OASCR and the Computational Materials Science Network of BES under
Contract No. DE-AC05-00OR22725 with UT-Battelle LLC, and also by NSF DMR-0341874.
\end{acknowledgments}

\bibliographystyle{apsrev}
\bibliography{Jdelta_dpl}

\begin{thebibliography}{28}
\expandafter\ifx\csname natexlab\endcsname\relax\def\natexlab#1{#1}\fi
\expandafter\ifx\csname bibnamefont\endcsname\relax
  \def\bibnamefont#1{#1}\fi
\expandafter\ifx\csname bibfnamefont\endcsname\relax
  \def\bibfnamefont#1{#1}\fi
\expandafter\ifx\csname citenamefont\endcsname\relax
  \def\citenamefont#1{#1}\fi
\expandafter\ifx\csname url\endcsname\relax
  \def\url#1{\texttt{#1}}\fi
\expandafter\ifx\csname urlprefix\endcsname\relax\def\urlprefix{URL }\fi
\providecommand{\bibinfo}[2]{#2}
\providecommand{\eprint}[2][]{\url{#2}}

\bibitem[{\citenamefont{Holtschneider et~al.}(2005)\citenamefont{Holtschneider,
  Selke, and Leidl}}]{Leidl05}
\bibinfo{author}{\bibfnamefont{M.}~\bibnamefont{Holtschneider}},
  \bibinfo{author}{\bibfnamefont{W.}~\bibnamefont{Selke}}, \bibnamefont{and}
  \bibinfo{author}{\bibfnamefont{R.}~\bibnamefont{Leidl}},
  \bibinfo{journal}{Phys. Rev. B} \textbf{\bibinfo{volume}{72}},
  \bibinfo{pages}{064443} (\bibinfo{year}{2005}).

\bibitem[{\citenamefont{Landau and Binder}(1981)}]{Landau81}
\bibinfo{author}{\bibfnamefont{D.~P.} \bibnamefont{Landau}} \bibnamefont{and}
  \bibinfo{author}{\bibfnamefont{K.}~\bibnamefont{Binder}},
  \bibinfo{journal}{Phys. Rev. B} \textbf{\bibinfo{volume}{24}},
  \bibinfo{pages}{1391} (\bibinfo{year}{1981}).

\bibitem[{\citenamefont{Pelcovits and Nelson}(1976)}]{Pelcovits76}
\bibinfo{author}{\bibfnamefont{R.~A.} \bibnamefont{Pelcovits}}
  \bibnamefont{and} \bibinfo{author}{\bibfnamefont{D.~R.}
  \bibnamefont{Nelson}}, \bibinfo{journal}{Phys. Letts.}
  \textbf{\bibinfo{volume}{57A}}, \bibinfo{pages}{23} (\bibinfo{year}{1976}).

\bibitem[{\citenamefont{Nelson and Pelcovits}(1977)}]{Nelson77}
\bibinfo{author}{\bibfnamefont{D.~R.} \bibnamefont{Nelson}} \bibnamefont{and}
  \bibinfo{author}{\bibfnamefont{R.~A.} \bibnamefont{Pelcovits}},
  \bibinfo{journal}{Phys. Rev. B} \textbf{\bibinfo{volume}{16}},
  \bibinfo{pages}{2191} (\bibinfo{year}{1977}).

\bibitem[{\citenamefont{Kosterlitz and Santos}(1978)}]{Kosterlitz78}
\bibinfo{author}{\bibfnamefont{J.~M.} \bibnamefont{Kosterlitz}}
  \bibnamefont{and} \bibinfo{author}{\bibfnamefont{M.~A.}
  \bibnamefont{Santos}}, \bibinfo{journal}{J. Phys. C}
  \textbf{\bibinfo{volume}{11}}, \bibinfo{pages}{2835} (\bibinfo{year}{1978}).

\bibitem[{\citenamefont{Cowley et~al.}(1993)\citenamefont{Cowley, Aharony,
  Birgeneau, Pelcovits, Shirane, and Thurston}}]{Cowley93}
\bibinfo{author}{\bibfnamefont{R.~A.} \bibnamefont{Cowley}},
  \bibinfo{author}{\bibfnamefont{A.}~\bibnamefont{Aharony}},
  \bibinfo{author}{\bibfnamefont{R.~J.} \bibnamefont{Birgeneau}},
  \bibinfo{author}{\bibfnamefont{R.~A.} \bibnamefont{Pelcovits}},
  \bibinfo{author}{\bibfnamefont{G.}~\bibnamefont{Shirane}}, \bibnamefont{and}
  \bibinfo{author}{\bibfnamefont{T.~R.} \bibnamefont{Thurston}},
  \bibinfo{journal}{Z. Phys. B} \textbf{\bibinfo{volume}{93}},
  \bibinfo{pages}{5} (\bibinfo{year}{1993}).

\bibitem[{\citenamefont{Kosterlitz et~al.}(1976)\citenamefont{Kosterlitz,
  Nelson, and Fisher}}]{Kosterlitz76}
\bibinfo{author}{\bibfnamefont{J.~M.} \bibnamefont{Kosterlitz}},
  \bibinfo{author}{\bibfnamefont{D.~R.} \bibnamefont{Nelson}},
  \bibnamefont{and} \bibinfo{author}{\bibfnamefont{M.~E.}
  \bibnamefont{Fisher}}, \bibinfo{journal}{Phys. Rev. B}
  \textbf{\bibinfo{volume}{13}}, \bibinfo{pages}{13} (\bibinfo{year}{1976}).

\bibitem[{\citenamefont{Landau and Binder}(1978)}]{Landau78}
\bibinfo{author}{\bibfnamefont{D.~P.} \bibnamefont{Landau}} \bibnamefont{and}
  \bibinfo{author}{\bibfnamefont{K.}~\bibnamefont{Binder}},
  \bibinfo{journal}{Phys. Rev. B} \textbf{\bibinfo{volume}{17}},
  \bibinfo{pages}{2328} (\bibinfo{year}{1978}).

\bibitem[{\citenamefont{Mermin and Wagner}(1966)}]{Mermin66}
\bibinfo{author}{\bibfnamefont{M.~E.} \bibnamefont{Mermin}} \bibnamefont{and}
  \bibinfo{author}{\bibfnamefont{H.}~\bibnamefont{Wagner}},
  \bibinfo{journal}{Phys. Rev. Lett.} \textbf{\bibinfo{volume}{17}},
  \bibinfo{pages}{1133} (\bibinfo{year}{1966}).

\bibitem[{\citenamefont{Polyakov}(1975)}]{Polyakov75}
\bibinfo{author}{\bibfnamefont{A.~M.} \bibnamefont{Polyakov}},
  \bibinfo{journal}{Phys. Letts.} \textbf{\bibinfo{volume}{59B}},
  \bibinfo{pages}{79} (\bibinfo{year}{1975}).

\bibitem[{\citenamefont{Br\'ezin and
  Zinn-Justin}(1976{\natexlab{a}})}]{Zinn-Justin76}
\bibinfo{author}{\bibfnamefont{E.}~\bibnamefont{Br\'ezin}} \bibnamefont{and}
  \bibinfo{author}{\bibfnamefont{J.}~\bibnamefont{Zinn-Justin}},
  \bibinfo{journal}{Phys. Rev. Lett.} \textbf{\bibinfo{volume}{36}},
  \bibinfo{pages}{691} (\bibinfo{year}{1976}{\natexlab{a}}).

\bibitem[{\citenamefont{Br\'ezin and
  Zinn-Justin}(1976{\natexlab{b}})}]{Brezin76}
\bibinfo{author}{\bibfnamefont{E.}~\bibnamefont{Br\'ezin}} \bibnamefont{and}
  \bibinfo{author}{\bibfnamefont{J.}~\bibnamefont{Zinn-Justin}},
  \bibinfo{journal}{Phys. Rev. B} \textbf{\bibinfo{volume}{14}},
  \bibinfo{pages}{3110} (\bibinfo{year}{1976}{\natexlab{b}}).

\bibitem[{\citenamefont{Aharony}(1978)}]{Aharony78}
\bibinfo{author}{\bibfnamefont{A.}~\bibnamefont{Aharony}},
  \bibinfo{journal}{Phys. Rev. B} \textbf{\bibinfo{volume}{18}},
  \bibinfo{pages}{3328} (\bibinfo{year}{1978}).

\bibitem[{\citenamefont{Hu}(2001)}]{Hu01}
\bibinfo{author}{\bibfnamefont{X.}~\bibnamefont{Hu}}, \bibinfo{journal}{Phys.
  Rev. Lett.} \textbf{\bibinfo{volume}{87}}, \bibinfo{pages}{057004}
  (\bibinfo{year}{2001}).

\bibitem[{\citenamefont{Miyatake et~al.}(1986)\citenamefont{Miyatake, Yamamoto,
  Kim, Toyonaga, and Nagai}}]{Miyatake86}
\bibinfo{author}{\bibfnamefont{Y.}~\bibnamefont{Miyatake}},
  \bibinfo{author}{\bibfnamefont{M.}~\bibnamefont{Yamamoto}},
  \bibinfo{author}{\bibfnamefont{J.~J.} \bibnamefont{Kim}},
  \bibinfo{author}{\bibfnamefont{M.}~\bibnamefont{Toyonaga}}, \bibnamefont{and}
  \bibinfo{author}{\bibfnamefont{O.}~\bibnamefont{Nagai}}, \bibinfo{journal}{J.
  Phys. C} \textbf{\bibinfo{volume}{19}}, \bibinfo{pages}{2539}
  (\bibinfo{year}{1986}).

\bibitem[{\citenamefont{Loison et~al.}(2004)\citenamefont{Loison, Qin, Schotte,
  and Jin}}]{Loison04}
\bibinfo{author}{\bibfnamefont{D.}~\bibnamefont{Loison}},
  \bibinfo{author}{\bibfnamefont{C.}~\bibnamefont{Qin}},
  \bibinfo{author}{\bibfnamefont{K.~D.} \bibnamefont{Schotte}},
  \bibnamefont{and} \bibinfo{author}{\bibfnamefont{X.~F.} \bibnamefont{Jin}},
  \bibinfo{journal}{Ero. Phys. J. B} \textbf{\bibinfo{volume}{41}},
  \bibinfo{pages}{395} (\bibinfo{year}{2004}).

\bibitem[{\citenamefont{Zhou et~al.}(2004)\citenamefont{Zhou, Kennett, Wan,
  Berciu, and Bhatt}}]{Zhou04}
\bibinfo{author}{\bibfnamefont{C.}~\bibnamefont{Zhou}},
  \bibinfo{author}{\bibfnamefont{M.~P.} \bibnamefont{Kennett}},
  \bibinfo{author}{\bibfnamefont{X.}~\bibnamefont{Wan}},
  \bibinfo{author}{\bibfnamefont{M.}~\bibnamefont{Berciu}}, \bibnamefont{and}
  \bibinfo{author}{\bibfnamefont{R.~N.} \bibnamefont{Bhatt}},
  \bibinfo{journal}{Phys. Rev. B} \textbf{\bibinfo{volume}{69}},
  \bibinfo{pages}{144419} (\bibinfo{year}{2004}).

\bibitem[{\citenamefont{Binder and Landau}(1984)}]{Binder84}
\bibinfo{author}{\bibfnamefont{K.}~\bibnamefont{Binder}} \bibnamefont{and}
  \bibinfo{author}{\bibfnamefont{D.~P.} \bibnamefont{Landau}},
  \bibinfo{journal}{Phys. Rev. B} \textbf{\bibinfo{volume}{30}},
  \bibinfo{pages}{1477} (\bibinfo{year}{1984}).

\bibitem[{\citenamefont{Fisher}(1975)}]{Fisher75}
\bibinfo{author}{\bibfnamefont{M.~E.} \bibnamefont{Fisher}},
  \bibinfo{journal}{Phys. Rev. Lett.} \textbf{\bibinfo{volume}{34}},
  \bibinfo{pages}{1634} (\bibinfo{year}{1975}).

\bibitem[{\citenamefont{Nelson and Rudnick}(1975)}]{Nelson75}
\bibinfo{author}{\bibfnamefont{D.~R.} \bibnamefont{Nelson}} \bibnamefont{and}
  \bibinfo{author}{\bibfnamefont{J.}~\bibnamefont{Rudnick}},
  \bibinfo{journal}{Phys. Rev. Lett.} \textbf{\bibinfo{volume}{35}},
  \bibinfo{pages}{178} (\bibinfo{year}{1975}).

\bibitem[{\citenamefont{Nelson}(1976)}]{Nelson76}
\bibinfo{author}{\bibfnamefont{D.~R.} \bibnamefont{Nelson}},
  \bibinfo{journal}{AIP Conf. Proc.} \textbf{\bibinfo{volume}{29}},
  \bibinfo{pages}{450} (\bibinfo{year}{1976}).

\bibitem[{\citenamefont{Rudnick and Nelson}(1976)}]{Rudnick76}
\bibinfo{author}{\bibfnamefont{J.}~\bibnamefont{Rudnick}} \bibnamefont{and}
  \bibinfo{author}{\bibfnamefont{D.~R.} \bibnamefont{Nelson}},
  \bibinfo{journal}{Phys. Rev. B} \textbf{\bibinfo{volume}{13}},
  \bibinfo{pages}{2208} (\bibinfo{year}{1976}).

\bibitem[{\citenamefont{Fisher and Nelson}(1974)}]{Nelson74}
\bibinfo{author}{\bibfnamefont{M.~E.} \bibnamefont{Fisher}} \bibnamefont{and}
  \bibinfo{author}{\bibfnamefont{D.~R.} \bibnamefont{Nelson}},
  \bibinfo{journal}{Phys. Rev. Lett.} \textbf{\bibinfo{volume}{32}},
  \bibinfo{pages}{1350} (\bibinfo{year}{1974}).

\bibitem[{\citenamefont{Fisher}(1974)}]{Fisher74}
\bibinfo{author}{\bibfnamefont{M.~E.} \bibnamefont{Fisher}},
  \bibinfo{journal}{Rev. Mod. Phys.} \textbf{\bibinfo{volume}{46}},
  \bibinfo{pages}{597} (\bibinfo{year}{1974}).

\bibitem[{\citenamefont{Kadanoff et~al.}(1976)\citenamefont{Kadanoff, Houghton,
  and Yalabik}}]{Kadanoff76}
\bibinfo{author}{\bibfnamefont{L.~P.} \bibnamefont{Kadanoff}},
  \bibinfo{author}{\bibfnamefont{A.}~\bibnamefont{Houghton}}, \bibnamefont{and}
  \bibinfo{author}{\bibfnamefont{M.~C.} \bibnamefont{Yalabik}},
  \bibinfo{journal}{J. Stat. Phys.} \textbf{\bibinfo{volume}{14}},
  \bibinfo{pages}{171} (\bibinfo{year}{1976}).

\bibitem[{\citenamefont{Kosterlitz and Thouless}(1973)}]{Thouless73}
\bibinfo{author}{\bibfnamefont{J.~M.} \bibnamefont{Kosterlitz}}
  \bibnamefont{and} \bibinfo{author}{\bibfnamefont{D.~J.}
  \bibnamefont{Thouless}}, \bibinfo{journal}{J. Phys. C}
  \textbf{\bibinfo{volume}{6}}, \bibinfo{pages}{1181} (\bibinfo{year}{1973}).

\bibitem[{\citenamefont{Peczak and Landau}(1991)}]{Peczak91}
\bibinfo{author}{\bibfnamefont{P.}~\bibnamefont{Peczak}} \bibnamefont{and}
  \bibinfo{author}{\bibfnamefont{D.~P.} \bibnamefont{Landau}},
  \bibinfo{journal}{Phys. Rev. B} \textbf{\bibinfo{volume}{43}},
  \bibinfo{pages}{1048} (\bibinfo{year}{1991}).

\bibitem[{\citenamefont{Rastelli et~al.}(2004)\citenamefont{Rastelli, Regina,
  and Tassi}}]{Rastelli04}
\bibinfo{author}{\bibfnamefont{E.}~\bibnamefont{Rastelli}},
  \bibinfo{author}{\bibfnamefont{S.}~\bibnamefont{Regina}}, \bibnamefont{and}
  \bibinfo{author}{\bibfnamefont{A.}~\bibnamefont{Tassi}},
  \bibinfo{journal}{Phys. Rev. B} \textbf{\bibinfo{volume}{70}},
  \bibinfo{pages}{174447} (\bibinfo{year}{2004}).

\end{thebibliography}
\end{document}